\documentclass[multphys,vecphys]{svmult}
\usepackage{amssymb}
\usepackage[mathscr]{eucal}
\usepackage{citesort}
\usepackage{url}
\usepackage{a4}
\usepackage{cite}
\usepackage{graphicx}
\usepackage{latexsym}
\makeindex
\begin{document}
\title*{Concepts of
 Hyperbolicity and Relativistic Continuum Mechanics}

\author{Robert Beig
\institute{Institut f\"ur Theoretische Physik der Universit\"at
der Universit\"at Wien, A-1090 Wien, Boltzmanngasse 5}
\texttt{robert.beig@univie.ac.at}} \maketitle

\begin{abstract}
After a short introduction to the characteristic geometry
underlying weakly hyperbolic systems of partial differential
equations we review the notion of symmetric hyperbolicity of
first-order systems and that of regular hyperbolicity of
second-order systems. Numerous examples are provided, mainly taken
from nonrelativistic and relativistic continuum mechanics.

\end{abstract}

\section{Introduction}\label{sec:1}
The notion of hyperbolicity of a partial differential equation
(PDE), or a system of PDE's, is central for the field theories of
mathematical physics. It is closely related to the well-posedness
of the Cauchy problem and to the causal structure underlying these
theories. In standard theories describing relativistic fields in
vacuo this causal structure is that given by the spacetime metric,
a second-order symmetric tensor of Lorentzian signature. If matter
is included, things become both more complicated and more subtle.
In fact, the awareness of some of those complications predates
Relativity by centuries. An example is afforded by the phenomenon,
already studied by Huygens, of birefringence in crystal optics
\footnote{For a fascinating account of the history of the
associated mathematics see \cite{Ga2}.}.

There is currently an increase of attention in the field of
Relativity, due in part to demands from Numerical Relativity,
devoted to certain notions of hyperbolicity applied to the
Einstein equations (for an excellent review see \cite{FR}). There
the main challenge, not discussed in the present notes at all,
comes from the fact that, already in vacuum, the Einstein
equations by themselves, i.e. prior to the imposition of any gauge
conditions, are not hyperbolic. The main burden, then, is to find
a ``hyperbolic reduction" turning the Einstein equations, or a
subset thereof, into a hyperbolic system appropriate for the
purpose at hand. However the complications in the causal structure
one finds in continuum mechanics, which are our main focus here,
are absent in the Einstein vacuum case - at least for the
reductions proposed so far. Of course, these complications do come
into play ultimately once matter-couplings are included.

These notes attempt an elementary introduction to some notions of
hyperbolicity and the ``characteristic geometry" associated with or
underlying these notions. The section following this one is
devoted to the general notion of a hyperbolic polynomial, which in
our case of course arises as the characteristic polynomial of a
PDE. It is interesting that this notion is on one hand restrictive
enough to encode essentially all the required features of a theory
in order to be ``causal" - on the other hand flexible enough to
account for an amazing variety of phenomena - relativistic or
nonrelativistic - ranging from gravitational radiation to water
waves or phonons in a crystal. We devote a significant fraction of
Sect.2 to examples, which at least in their nonrelativistic guise
all appear in the standard literature such as \cite{CH}, though
not perhaps from the unified viewpoint pursued here. Some of these
examples are not fully worked out, but perhaps the interested
reader is encouraged to fill in more details, possibly using some
of the cited literature. We hope that some workers in Relativity,
even if they have little interest in continuum mechanics for its
own sake, find these examples useful for their understanding of
the notion of hyperbolicity. While hyperbolicity of the
characteristic polynomial of a theory is important, it is not in
general sufficient for the well-posedness of the initial value
problem for that theory. Well-posedness is the subject of our
Sect.3. We recall the notion of a symmetric hyperbolic system of a
system of 1st order PDE's, which is indeed sufficient for
well-posedness. A similar role for 2nd order equations is played
by a class of systems, which were to some extent implicit in the
literature, and for which an elaborate theory has been recently
developed in \cite{Ch1,Ch2}. These systems are called regular
hyperbolic.  They encompass many second order systems arising in
physics one would like to qualify as being hyperbolic - such as
the Einstein equations in the harmonic gauge. If applicable, the
notion of regular hyperbolicity is particularly natural for
systems of 2nd order derivable from an action principle, as is the
case for many problems of continuum mechanics. We show the fact,
obvious for symmetric hyperbolic systems and easy-to-see although
not completely trivial for regular hyperbolic ones, that these
systems are special cases of weakly hyperbolic systems, i.e. ones
the determinant of whose principal symbol is a hyperbolic
polynomial. We also touch the question of whether a system of the
latter type can be reduced to one of the former type by increasing
the number of dependent variables. Throughout this section our
treatment will be informal in the sense of ignoring specific
differentiability requirements. We also do not touch questions of
global wellposedness.

\section{Hyperbolic Polynomials}\label{sec:2}

The PDE's we are interested in are of the form
\begin{equation}\label{equgen}
M^{\mu_1\ldots\mu_l}_{AB}(x,f,\partial f,\ldots,\partial^{(l-1)}
f)\partial_{\mu_1}\dots\partial_{\mu_l}f^B + \textrm{lower order terms}.
\end{equation}
Here $A,B=1,...m$ and $\mu_i = 1,\ldots n$. Relevant equations of
this form are the Euler equations for a barotropic fluid (for
$n=4, l=1, m=4$), the Einstein equations (for $n=4, l=2, m=10$) or
the equations governing an ideal elastic solid (for $n=4, l=2,
m=3$). The Maxwell equations, in the form they are originally written down,
are not of this form, but a suitable subset of them is, as we will
discuss later.

The principal symbol of the PDE (\ref{equgen}) is defined as
\begin{equation} \label{sym}
M_{AB}(k)= M^{\mu_1 \dots\mu_l}_{AB}k_{\mu_1}\dots k_{\mu_l},\hspace{0.3cm} k_\mu \in
(\mathbb{R}^n)^*
\end{equation}
We here suppress the dependence on $x$ and on $f$.
 The characteristic polynomial $P(k)$ is defined by
$P(k)=det\hspace{0.1cm}M_{AB}(k)$, where the determinant is taken with respect to some volume form
on $f$-space $\subset \mathbb{R}^m$. The polynomial $P(k)$ is homogenous of degree $p=m\cdot l$. A
homogenous polynomial of degree $p>0$ is called hyperbolic with respect to $\xi_\mu \in(\mathbb
R^n)^*$ if $P(\xi) \neq 0$ and the map $\lambda \mapsto P(\eta + \lambda \xi)$, itself a polynomial
of degree
$p$, has only real roots $\lambda_i, i=1,\cdots p$ for all $\eta \in (\mathbb R^n)^*$.\\
The roots $\lambda_i(\xi,\eta)$ need not be distinct. If, for all $\eta$ with $\eta \wedge \xi \neq
0$, $\lambda_i(\xi,\eta) \neq \lambda_j(\xi,\eta)$ for $i \neq j$, $P$ is called strictly
hyperbolic \footnote{This case is not general enough for the purposes of physics. Furthermore
there exist physically relevant cases of non-strictly hyperbolic polynomials which are stable,
in the sense that they possess open neighbourhoods in the set of hyperbolic polynomials
just containing non-strictly hyperbolic ones~\cite{J,Hoe}.}). We write $\mathscr{C}^*$ for the set of $k \in
(\mathbb{R}^n)^*\setminus\{ 0 \}$, where $P$ vanishes. It is sometimes called the cone of characteristic conormals.\\
 It is clear that a product of hyperbolic polynomials is
hyperbolic. Also, if a hyperbolic polynomial can be factorized into polynomials of lower degree (in
which case it is called reducible), these factors are also hyperbolic. There is a wealth of
information which can be inferred about a polynomial $P(k)$ if it is hyperbolic. Before explaining
some of this, we look at a few examples for hyperbolic polynomials.

\bigskip
{\bf{Example 1}}: $P(k) = (a,k) = a^\mu k_\mu$ for some nonzero
$a^\mu \in \mathbb{R}^n$. The set $\mathscr{C}^*$ is a punctured hyperplane
$\subset (\mathbb{R}^n)^*$.\\
Clearly $P(k)$ is hyperbolic with respect to any $\xi_\mu$ such that
$a^\mu \xi_\mu \neq 0$. The polynomial
$P(k)= (a_1,k)(a_2,k)(a_3,k)$, with $a_1, a_2, a_3$ linearly
independent $\in \mathbb{R}^3$, arises in the problem of finding,
for a three dimensional positive definite metric, a coordinate
system in which the metric is diagonal (see \cite{DY}) - which
shows that hyperbolic problems can also arise in purely Riemannian
contexts.

\bigskip
{\bf{Example 2}}: $P(k)=\gamma^{\mu\nu}k_\mu k_\nu$, where
$\gamma^{\mu\nu}$ is a (contravariant) metric of Lorentzian
signature $(-,+,\ldots,+)$. The set $\mathscr{C}^*$ is the two-sheeted
Minkowski light cone.\\
 When $n=2$, $P(k)$ is hyperbolic
with respect to any non-null $\xi_\mu$, when $n>2$, $P(k)$ is hyperbolic
with respect to any $\xi_\mu$ with $\gamma^{\mu\nu} \xi_\mu \xi_\nu < 0$,
i.e. $\xi$ is timelike with respect to $\gamma^{\mu\nu}$. Checking that
$P(k)$ is hyperbolic according to our definition is equivalent to
the so-called reverse Cauchy-Schwarz inequality for two covectors
one of which is timelike or null with respect to $\gamma^{\mu\nu}$ (which
is the mathematical rationale behind the the twin ``paradox" of
Relativity). Surprisingly there are similar inequalities for general
hyperbolic polynomials (see \cite{Ga1}) which play a role in
diverse fields of mathematics \cite{BGLS}.

Example 2 is of course the most familiar one. If it arises from nonrelativistic field theory, the
quantity $\gamma^{\mu\nu}$ currently runs under the name of the ``Unruh or acoustic metric"
\cite{BLV} (see also \cite{Chr}) in the Relativity community. It is not an elementary object
of the theory, but is built as follows: Take first the Galilean metric $h^{\mu\nu}$, a symmetric tensor with signature
$(0,+,\ldots,+)$ together with a nonzero covector $\tau_\mu$ satisfying $h^{\mu\nu} \tau_\nu =0$:
these are the absolute elements . Then pick a 4-vector $u^\mu$ normalized so that $u^\mu \tau_\mu
=1$ and define $\gamma^{\mu\nu} = h^{\mu\nu} - c^{-2} u^\mu u^\nu$. This describes waves
propagating isotropically at phase velocity $c$ in the rest system, defined by $u^\mu$, of a
material medium. The relativistic version of the above is as follows: Start with the spacetime
metric $g^{\mu\nu}$ and define $\gamma^{\mu\nu} = g^{\mu\nu} + (1 - c^{-2})u^\mu u^\nu$, where
$u^\mu$ is normalized by taking $\tau_\mu = - g_{\mu\nu} u^\nu$, with $g_{\mu\nu}$ the covariant
spacetime metric defined by $g_{\mu\nu} g^{\nu\lambda}= \delta_\mu{}^\lambda$. Note: if there are
metrics $\gamma_1^{\mu\nu}$,$\gamma_2^{\mu\nu}$ with $c_2 < c_1$, then the
``faster" cone lies inside the slower one. We will come back to this
point later.

\bigskip
{\bf{Example 3}}: $P(k) = s^{\mu\nu}k_\mu k_\nu$, where
$s^{\mu\nu}$ has signature $(-,+,\ldots,+,0,\ldots,0)$, is hyperbolic
with respect to any $\xi$ such that $s^{\mu\nu}\xi_\mu \xi_\nu <0$.\\
Here is a case occurring in the real world. Let $g_{\mu\nu}$ be a Lorentz metric on $\mathbb{R}^4$,
$u^\mu$ a normalized timelike vector, i.e. $g_{\mu\nu}u^\mu u^\nu =-1$, $F_{\mu\nu}=F_{[\mu\nu]}$
nonzero with $F_{\mu\nu} u^\nu =0$. The quadratic form $s^{\mu\nu}= - e u^\mu u^\nu + 1/2\
F_{\rho\sigma} F^{\rho\sigma}g^{\mu\nu} - F^\mu{}_\rho F^{\nu\rho}$, with $e>0$, has signature
$(-,+,+,0)$. The characteristic cone $\mathscr{C}^*$ of $P(k)=0$ consists of two
hyperplanes punctured at the origin. When $e$ is interpreted as
$e=$``energy density + pressure" and
$F_{\mu\nu}$ as the frozen-in magnetic field of an ideally conducting plasma, then $P(k)$
describes the Alfven modes of relativistic magnetohydrodynamics \cite{Z}\cite{Ani}.

\bigskip
{\bf{Example 4}}: Let $n=4$, $\varepsilon_{\mu\nu\lambda\rho}$
some volume form on $\mathbb{R}^4$ and $m^{\mu\nu\lambda\rho}=
m^{[\mu\nu][\lambda\rho]}$. With $G^{\mu\nu\rho\sigma} =
\varepsilon_{\alpha\beta\delta\epsilon}\hspace{0.1cm}\varepsilon_{\kappa\phi\psi\omega}
\hspace{0.1cm}m^{\alpha\beta\kappa(\mu}
m^{\nu|\delta\phi|\rho} m^{\sigma)\epsilon\psi\omega}$ we
define $P$ by $P(k) = G^{\mu\nu\rho\sigma} k_\mu k_\nu k_\rho
k_\sigma$.\\
As a special case take $m^{\mu\nu\lambda\rho}$ of the form
$m^{\mu\nu\lambda\rho}= h^{\lambda[\mu} h^{\nu]\rho} -
e^{\lambda[\mu} u^{\nu]} u^\rho + e^{\rho[\mu} u^{\nu]}
u^\lambda$, where the symmetric tensors $h^{\mu\nu}$ and
$s^{\mu\nu}$, both of signature $(0,+++)$, satisfy $h^{\mu\nu}
\tau_\nu = e^{\mu\nu} \tau_\nu = 0$ for $u^\mu \tau_\mu \neq 0$:
this is the situation encountered in crystal optics with the
nonzero eigenvalues of $e^{\mu\nu}$ relative to $h^{\mu\nu}$ being
essentially the dielectric constants. The crystal is optically biaxial
or triaxial, depending on the number of mutually different eigenvalues.
The 4th order polynomial
$P(k)$ turns out to be hyperbolic with respect to all $\xi_\mu$ in some
neighbourhood of $\xi_\mu = \tau_\mu$ , and the associated
characteristic cone is the Fresnel surface (see e.g. \cite{He}).
For an optically isotropic medium or in vacuo $P(k)$ is reducible,
in fact the square of a quadratic polynomial of the type of
Example 2. We leave the details as an exercise. More general
conditions on $m^{\mu\nu\lambda\rho}$
in order for $P(k)$ to be hyperbolic can be inferred from \cite{LH}.\\
The quartic polynomial $P(k)$, as defined above, comes from a
generalized (``pre-metric") version of electrodynamics (see
\cite{HO}), as follows: Let $F_{\mu\nu}$ be the electromagnetic
field strength and write $H^{\mu\nu} = m^{\mu\nu\lambda\rho}
F_{\lambda\rho}$ for the electromagnetic excitation. The premetric
Maxwell equations then take the form
\begin{equation}\label{max1}
\partial_{[\mu} (\varepsilon_{\nu\lambda]\rho\sigma}H^{\rho\sigma}) =
J_{\mu\nu\lambda},\hspace{0.5cm} \partial_{[\mu}F_{\nu\lambda]}=0,
\end{equation}
where $J_{\mu\nu\lambda}$ is the charge three form \footnote{These equations play a certain role in
current searches for violations of Lorentz invariance in electrodynamics \cite{KM}}. The Eq.'s
(\ref{max1}) reduce to the standard ones in vacuo when $m^{\mu\nu\lambda\rho} \sim
g^{\lambda[\mu}g^{\nu]\rho}$ with $g^{\mu\nu}$ the metric of spacetime. If one sets
$m^{\mu\nu\lambda\rho} = \gamma^{\lambda[\mu}\gamma^{\nu]\rho}$, with $\gamma^{\mu\nu}= h^{\mu\nu}
- c^{-1}u^\mu u^\nu$, $h^{\mu\nu}$ the Galilean metric and $u^\mu$ a constant vector field s.th.
$u^\mu \tau_\mu = 1$, one has the Maxwell equations in a ``Galilean" (not Galilean-invariant)
version with $u^\mu$ describing the rest system of the aether (see \cite{Tr}). One then looks at
hypersurfaces along which singularities can propagate. The result is that the conormal $n_\mu$ of
such surfaces has to satisfy $P(n)=0$. Put differently, one can look at the 8 x 6 - principal
symbol of the Maxwell equations: then $P(k)=0$ is exactly the condition for this principal symbol
to have nontrivial kernel. If one considered an appropriately chosen subset amongst (\ref{max1}),
the evolution equations, one would obtain an equation of the form (\ref{equgen}), whose
characteristic polynomial contains $P(k)$ as a factor. We will treat the vacuum case of this later.

\bigskip
Our last and most complicated example comes from elasticity
\cite{BS}, namely \\
{\bf{Example 5}}: Take $n=4,l=2,m=3$ in Eq.(\ref{equgen}) with
\begin{equation} \label{el}
M_{AB}^{\mu\nu}= - G_{AB}u^\mu u^\nu + C_{AB}^{\mu\nu},
\end{equation}
where $G_{AB}=G_{(AB)}$ and $C_{AB}^{\mu\nu} = C_{BA}^{\nu\mu}$ and $C_{AB}^{\mu\nu}\tau_\nu =0$
for some covector $\tau$ satisfying $(u,\tau)=u^\mu \tau_\mu =1$. The theory is intrinsically
quasilinear: all quantities entering Eq.(\ref{el}) are functions of $f$ and $\partial f$ and in
general also of $x$. For example $f^A$ is required to have maximal rank, and $u^\mu$ satisfies
$u^\mu (\partial_\mu f^A)=0$. Furthermore $C^{\mu\nu}_{AB}= C_{ADBE}(\partial_\rho
f^D)(\partial_\sigma f^E)h^{\rho\mu} h^{\sigma\nu}$, with $h^{\mu\nu}, \tau_\mu$ being, in the
nonrelativistic case, the absolute Galilean objects, or, in the relativistic case, $h^{\mu\nu} =
g^{\mu\nu} + u^\mu u^\nu$ and $\tau_\mu = - g_{\mu\nu}u^\nu$.

There are the following basic constitutive assumptions
\begin{equation}\label{con}
G_{AB} \hspace{0.2cm}\textrm{is positive definite} \hspace{1cm}
C_{AB}^{\mu\nu} m^A m^B \eta_\mu \eta_\nu > 0 \hspace{0.2cm}
\textrm{for}\hspace{0.2cm}m\neq0,\ \eta \wedge \tau \neq0
\end{equation}
Defining the linear map $({\bf{M}})^A{}_B$ by $({\bf{M}})^A{}_B (k) = - (u,k)^2 \: \delta^A{}_B +
G^{AD} C^{\mu\nu}_{DB} k_\mu k_\nu$, the polynomial $P(k)$ can, by general linear algebra, be
written as
\begin{equation} \label{elpol}
6\ P(k)= (tr{\bf{M}})^3 - 3\ (tr{\bf{M^2}}) (tr{\bf{M}}) + 2\ tr
{\bf{M^3}}.
\end{equation}
It will follow from a more general result, to be shown below, that the 6th order
polynomial $P(k)$ is hyperbolic
with respect to $\xi_\mu$ in some neighbourhood of $\tau_\mu$. In the special case of an isotropic solid,
the ``elasticity tensor" $C_{ABDE}$ has to be of the form
\begin{equation} \label{iso}
C_{ABDE} = l\ G_{AB}G_{DE} + 2 m\ G_{D(A}G_{B)E},
\end{equation}
and the second of Eq.(\ref{con}) is satisfied iff $c_2^2=m>0$,
$c_1^2=l + 2m>0$. The polynomial $P(k)$ turns out to reduce to the
form
\begin{equation}\label{iso1}
P(k) \sim (\gamma_1^{\mu\nu}k_\mu k_\nu)(\gamma_2^{\rho\sigma}k_\rho
k_\sigma)^2
\end{equation}
with $\gamma_{1,2}^{\mu\nu} = h^{\mu\nu} - c_{1,2}^{-2}\ u^\mu
u^\nu$. The quantities $c_1$ and $c_2$ are the phase velocities
of pressure and shear waves respectively. If the medium is elastically
anisotropic, such as a crystal, one can start by classifying
possible fourth-rank tensors $C_{ABDE}$ according to the symmetry
group of the crystal lattice, allow for dislocations, etc. The
richness of possible structure of $\mathscr{C}^*$ and the
corresponding range of captured physical phenomena - studied by
theoreticians and experimentalists - is enormous.

\bigskip
This ends our list of examples. We now turn to some general properties of hyperbolic polynomials
and their physical interpretation. It is clear from the definition that $\mathscr{C}^*$ has
codimension 1: since $P(\eta + \lambda \xi)$ has to have at least one complex root for each $\eta$,
and the roots are all real, there is at least one real root. And since there are no more than
$p\cdot l$ different roots, $\mathscr{C}^*$ can not have larger codimension. It is then known from real
algebraic geometry that $\mathscr{C}^*$ consists of smooth hypersurfaces outside a set of at least
codimension $2$ (see \cite{BR}). The roots $\lambda_i(\xi,\eta)$ can for fixed $\xi$ be assumed to
be ordered according to $\lambda_1 \leq \lambda_2 \leq \dots \leq \lambda_p$ for all $\eta$. The
set of points $k=\eta + \lambda_i(\xi,\eta)\xi$ is called the $i$'th sheet of $\mathscr{C}^*$. The
hypersurface $\mathscr{C}^*$ has to be smooth at all points $k$ lying on a line intersecting $p$ different
sheets \footnote{The reason is that a polynomial of order $p$ in one real variable, if it has $p$
different zeros, has non-vanishing derivative at each zero, so $k$ is a non-critical point of
$P$.}. In particular all sheets are everywhere smooth when $P$ is strictly
hyperbolic.\\
Next recall that the defining property of a hyperbolic polynomial refers to a particular covector
$\xi$. That covector however is not unique. It is contained in a unique connected, open, convex,
positive cone $\Gamma^*(\xi)$ of covectors $\xi'$ sharing with $\xi$ the property that $P(\xi')
\neq 0$ and $P(\eta + \lambda \xi')$ has only real zeros $\lambda_i (\xi',\eta)$ \cite{Ga1}. Note
that $\Gamma^*(\xi) = - \Gamma^*(-\xi)$. Furthermore $\partial \Gamma^*(\xi) \subset
\mathscr{C}^*$, and $\Gamma^*(\xi)$ is that connected component of the complement of
$\mathscr{C}^*$ containing $\xi$. Not all points of $\partial \Gamma^*(\xi)$ have to be smooth
points of $\mathscr{C}^*$.

The roots $\lambda_i(\xi,\eta), i=1,\dots p$, due to the homogeneity of $P$, are homogenous in
$\xi$ of order $-1$ and positively homogenous of order 1 as a function of $\eta$. They also satisfy
$\lambda_i (\xi,\eta) = - \lambda_{p+1-i}(\xi,\eta)$. At regular points of $\mathscr{C}^*$, i.e.
when the gradient of $P$ at $\eta + \lambda_i \xi$ is non-zero, $\lambda_i(\xi,\eta)$
is a smooth function of its arguments due to the implicit function theorem.\\
Next choose a vector $X \in
\mathbb{R}^n$ so that $(X,\xi')>0$ for all $\xi' \in
\Gamma^*(\xi)$. We call such a vector ``causal". We now look
at the intersection $S$ of the hyperplane $(X,\xi')=1$ with
$\mathscr{C}^*$. Note $(X,\xi')=1$ is transversal to
$\mathscr{C}^*$ at smooth points of $\mathscr{C}^*$, so $S$ is smooth there also.
Note also that $S$ may be empty, as in Ex.1.
 For reasons explained below, $S$ is often called
``slowness surface". To describe $S$ more concretely, we pick some $\tau \in \Gamma^*(\xi)$ with
$(X,\tau)=1$. The pair $X,\tau$ constitutes a ``rest frame". Using it, we can decompose every
covector $k$ as $k_\mu = \tau_\mu + k^\perp_\mu$ where $k^\perp$ is tangential to the hyperplane,
i.e. $(X,k^\perp)=0$. This $k$ lies on $\mathscr{C}^*$ iff $\lambda_i(\tau,k^\perp)=1$ for some
$i$. Thus $S$ consists of the sheets $\lambda_J(\tau,k^\perp)=1$, viewed as $(n-2)$-surfaces in
$k^\perp \in \mathbb{R}^{n-1}$. Here $J$ runs through some subset of the $i$'s parametrizing
$\lambda_i$ from above. Clearly, as $J$ increases, these sheets form a nested family of not
necessarily compact surfaces. \footnote{In particular, when sheets seem to pass through each other,
the two sides are counted as belonging to different sheets.} The innermost of these surfaces is
nothing but the intersection of $\partial \Gamma^*(\xi)$ with the hyperplane $(X,\xi')=1$ and is
hence convex. We call $\mathscr{C}^*(\xi)$ those components of $\mathscr{C}^*$ which consist of
half-rays connecting the origin with the points of $S$. In the examples 2, 4, 5 the set $S$ and
$\mathscr{C}^*(\xi)$ consist of at most 1, respectively 2 and 3 sheets. For the last-mentioned case, see
\cite{Du}. Not all the occurring sheets are compact. It is possible for example for $P(k)$ to be an
irreducible hyperbolic polynomial with some sheets of $S$ compact and others non-compact: this is
the case e.g. for the acoustic modes
in magnetohydrodynamics \cite{CH}.\\

We now explain the name ``slowness surface". Consider the hyperplane in $\mathbb{R}^n$ given by
$(x,k)=0$ for fixed $k \in \mathscr{C}^*$, i.e. the wave front of the plane wave associated with
$k$. To measure the ``speed" at which this wave front moves, decompose observers with tangent $V$
which ``move with this wave front", i.e. such that $(V,k)=0$, according to $V=X+v^\perp$. It follows
that $(v^\perp,k^\perp)=-1$. Thus, if there is a natural ``spatial" metric $h$ mapping elements
$l^\perp$ into elements $w^\perp = h \circ l^\perp$ orthogonal to $\tau$, one can define the ``phase
velocity" $v^\perp_{\mathrm{ph}}= - \Vert k^\perp \Vert^{-2}\  h \circ k^\perp$. Thus, the smaller
$k^\perp$ is, the larger the phase velocity. Of course the equation $(v^\perp,k^\perp)=-1$ does not
define $v^\perp$ uniquely. But there is a ``correct" choice for $V$ tangential to the wave front,
called ``ray or group velocity", which is independent of any spatial metric, and which is defined at
least when $k$ is a smooth point of $\mathscr{C}^*$: this $V$ is given by the conormal to
$\mathscr{C}^* \subset (\mathbb{R}^n)^*$ at $k$, which by duality is a vector $\in \mathbb{R}^n$.
If $k$ is in addition a non-critical point, this ray velocity $V^\mu$ is $\sim
\partial/\partial k_\mu P(k)$, which satisfies $k_\mu V^\mu =0$ by
the positive homogeneity of $P(k)$. The spatial group velocity in
the frame $X,\tau$ can then be written as
\begin{equation}\label{group}
(v_{\mathrm{gr}}^\perp)^\mu (k^\perp) = \bigg(\tau_\lambda \frac{\partial P}{\partial k_\lambda}\bigg)^{-1}\
\big(\delta^\mu{}_\nu - X^\mu \tau_\nu\big)\ \frac{\partial P}{\partial k_\nu}\bigg|_{k=\tau + k^\perp},
\end{equation}
which is also the textbook expression.\\
We should add a cautionary remark here. Although the differential topology of the slowness surface
is independent of the choice of $X$ satisfying $(X,\xi')>0$ for all $\xi' \in \Gamma^*(\xi)$, its
detailed appearance, and physical quantities such as phase velocity, group velocity or angle
between two rays do of course depend on the choice of rest system $X, \tau$ and a notion of spatial
metric with respect to that observer. Of course there will be, for any particular physical theory, a
singled-out class of rest systems, e.g. $\tau_\mu$ can be the absolute object in a Galilean
spacetime or be of the form $\tau_\mu = - g_{\mu\nu} X^\nu$ in a relativistic theory. Or the
slowness surface can have more symmetry (say symmetry with respect to reflection at the origin) in some
rest system than in others, as is the case with crystal optics or elasticity. For a careful
discussion of these issues, in the more
specialized context of ``ray-optical structures" on a Lorentzian spacetime, consult \cite{Pe}.\\
We now come back to the ``ray" concept. If $k$ is a smooth critical point of $\mathscr{C}^*$,
finding the map $k \mapsto V(k)$ is already a nontrivial problem in algebraic geometry \cite{R}. If
$k$ is not a smooth point of $\mathscr{C}^*$, there is no unique assignment of a group velocity to
$k$. Still well-defined is the set $\mathscr{C}$ of all $V \neq 0$ satisfying
\begin{equation}\label{dual}
(V,k)=0\hspace{0.7cm}\textrm{where}\hspace{0.4cm}P(k)=0,
\end{equation}
called the dual or ray cone. Loosely speaking, each sheet of the ray cone corresponds to a
spherical wave front tangent to (or ``supported by") the planar wave fronts defined by the different
points $k$ in some corresponding sheet of $\mathscr{C}^*$ \cite{CH}. There holds $(\mathscr{C}^*)^*
= \mathscr{C}$. The dual cone is again an algebraic cone, which,
 except in degenerate cases, is again the zero-set of a single
homogenous polynomial. The structure of this dual cone, in particular its singularity structure
which can be very complicated, is another difficult matter of real algebraic geometry. For example
the degree of its defining polynomial is in general much higher than that of $\mathscr{C}^*$ (see
\cite{S},\cite{GKZ}). This ``dual" polynomial need not be hyperbolic: in order to be hyperbolic it
would have to have a central sheet which is convex, which is not the case for some of the examples
one finds in the literature. In our examples from above the situation is as follows: In our Ex.1
the dual cone $\mathscr{C}^*$ consist of the two half-lines $\{ \alpha a^\mu | \alpha>0 \}$ and $\{
\alpha a^\mu|\alpha<0 \}$. The cone dual to the quadratic cone $g^{\mu \nu}k_\mu k_\nu=0$ in Ex.2
is given by $g_{\mu\nu}V^\mu V^\nu =0$ with $g_{\mu\nu} g^{\nu\lambda} = \delta_\mu{}^\lambda$. For
a nonrelativistic acoustic cone $\gamma^{\mu\nu} = h^{\mu\nu} - (1/c^2)u^\mu u^\nu$ we obtain for
the ray cone $\gamma_{\mu\nu} = h_{\mu\nu} - c^2 \tau_\mu \tau_\nu$, where $h_{\mu\nu}$ is the
unique tensor defined by $h_{\mu\nu} u^\nu = 0$ and $h_{\mu\nu} h^{\nu\lambda} =
\delta_\mu{}^\lambda - \tau_\mu u^\lambda$. If one has two sound cones, as in isotropic elasticity,
it is the faster ray cone which lies outside. In Ex.3 $\mathscr{C}$ is given as a subset of vectors
$X^\mu$ in a linear space $T$, which is the annihilator of the null space of $s^{\mu\nu}$, namely
where this subset is given by $s_{\mu\nu}X^\mu X^\nu =0$, where $s_{\mu\nu}$ is the inverse of
$s^{\mu\nu}$ on $T$. In the magnetohydrodynamic example the preceding statement corresponds to the
fact that Alfven waves ``travel along the direction of the magnetic field". For Ex.4 the ray cone
$\mathscr{C}$ is a 4th order cone of the same type as $\mathscr{C}^*$, a fact already known by
Amp\`ere in the case of crystal optics and shown generally in \cite{Ru}. For anisotropic elasticity
the structure of the ray cone does not seem to be fully known, except for a general upper bound on
its degree, namely 150 on grounds of general algebraic geometry (see \cite{Du},\cite{S},\cite{GKZ})
and detailed studies for certain specific crystal symmetries - which give rise to a beautiful
variety of acoustic phenomena \cite{Wo}\footnote{There are computer codes designed for algebraic
elimination, which might be worth applying to this problem \cite{GPS}}.

\section{Initial Value Problem}\label{sec:4}

We now come to the issue of posing an initial value problem for hyperbolic equations of the form of
Eq.(\ref{equgen}). This requires two things: firstly a notion of ``spacelike" initial value surface,
secondly a notion of domain of dependance. Not surprisingly these notions can be formulated purely
in terms of the characteristic polynomial. A hypersurface $\Sigma$ in $\mathbb{R}^n$ will be called
spacelike, if it has a conormal $n_\mu$ lying everywhere in $\Gamma^*(\xi)$ for some $\xi$. If the
equation (\ref{equgen}) is nonlinear, every property concerning the characteristic polynomial has
to refer to the data of some reference field $f_0$, i.e. the value of $f_0$ on $\Sigma$ and those
of its derivatives up to order $l-1$. It is then the case that $\Sigma$ is spacelike also for any
sufficiently near-by data. The reason is that $\xi' \in \Gamma^*(\xi)$ can be characterized by
$\lambda_1(\xi,\xi')>0$, and the eigenvalues $\lambda_i$, being zeros of a polynomial having real
roots only, depend continuously on the coefficients of this polynomial \cite{AKLM}. A point $x$ in
$\mathbb{R}^n$ is said to lie in the domain of dependence of $\Sigma$ if each causal curve (i.e.
each curve whose tangent vector $X$ satisfies $(X,\xi')\neq 0$ for all $\xi' \in \Gamma^*(\xi)$)
through $x$ which is inextendible intersects $\Sigma$ exactly once. The Cauchy problem for
Eq.(\ref{equgen}) is said to be well-posed if, for the above data, there is a unique solution in
some domain of dependence of $\Sigma$ and, secondly, if this solution depends in some appropriate
sense continuously on the data. The question then is whether well-posedness holds under the above
conditions. The answer is affirmative when Eq.(\ref{equgen}) is linear with constant coefficients
and the lower-order terms are absent. Then the initial value problem can be solved ``explicitly" by
using a  fundamental solution (``Green function" in the physics literature) - which in turn can be
obtained e.g. by the Fourier transform. By a refined version of a well-known argument in physics
texts concerning the wave equation in Minkowski space (see e.g. \cite{ABG}), one can show that the
fundamental solution is supported in $\Gamma(\xi)$, which is the closure of the set of causal
vectors just described. The set $\Gamma(\xi)$ is a closed, convex cone, dual to $\Gamma^*(\xi)$. If
the outermost component of the cone $\mathscr{C}(\xi)$ dual to $\mathscr{C}^*(\xi)$ is convex, its
closure is the same as $\Gamma(\xi)$, otherwise its convex closure is the same as $\Gamma(\xi)$. If
one is interested in finer details than just wellposedness, even the linear, constant-coefficient
case becomes very nontrivial. An example is the question of the existence
of ``lacunas", i.e.
regions in $\Gamma(\xi)$ where the fundamental solution vanishes. For isotropic elasticity
mentioned in Ex.4, when $c_2 < c_1$ (which is the experimentally relevant case), the fundamental
solution vanishes inside the inner shear cone determined by $c_2$. (Note that
``inner" and ``outer"
are interchanged under transition between normal and ray cone.) For anisotropic elasticity this
issue, or the somewhat related question of the detailed time decay, already presents great
difficulties (see \cite{Du,St}). The existence of lacunas for general, linear hyperbolic systems
with constant coefficients was studied in \cite{ABG}.

The problem now is that many field equations in physics give rise to variable coefficients, to
various forms of lower-order terms and-or nonlinearities. But if one has a system of PDE's with
hyperbolic characteristic polynomial (such systems are often called ``weakly hyperbolic"), which in
addition has a well posed initial value problem, a perturbation of the coefficients will in general
destroy the latter property  (see e.g. \cite{KO}). It is thus hard to get any further without
additional assumptions. One such assumption is that of having a symmetric hyperbolic system. This
is given by a system of the form of Eq.(\ref{equgen}) with $l=1$. It is furthermore assumed that
\begin{equation}\label{sym1}
M_{AB}^{\ \mu} = M_{(AB)}^{\ \mu}
\end{equation}
and that there exists $\xi_\mu$ so that
\begin{equation}\label{pos}
M_{AB}(\xi) = M_{AB}^{\ \mu}\ \xi_\mu \hspace{0.4cm} \textrm{is positive definite}.
\end{equation}
The symmetric hyperbolic system has a characteristic polynomial which is hyperbolic with respect to $\xi$.
To see this one simply observes that the equation
\begin{equation}\label{sympol}
det(M_{AB}^{\ \mu}(\eta_\mu + \lambda \xi_\mu)) = 0
\end{equation}
characterizes eigenvalues of the quadratic form $M_{AB}(\eta)$ relative to the metric $M_{AB}(\xi)$ -
and these eigenvalues have to be real. There is then, for quasilinear symmetric hyperbolic systems,
a rigorous existence statement
along the lines informally outlined at the beginning of this section \cite{Ka}.
The uniqueness part uses the concept of ``lens-shaped domains" (see e.g. \cite{FR}) which is essentially
equivalent to that of domain of dependence above.\\
Several field theories of physical importance naturally give rise to a symmetric hyperbolic system.
An example is afforded by the hydrodynamics of a perfect fluid both nonrelativistically and
relativistically \footnote{For an elegant treatment of the latter, see \cite{Fr}}. The most
prominent examples are perhaps the Maxwell equations in vacuo and the vacuum Bianchi identities in
the Einstein theory. For the latter this was first observed in \cite{Fri}. For completeness we
outline a proof for the well-known Maxwell case following \cite{Z}. We have that
\begin{equation}\label{max}
\nabla^\nu F_{\mu\nu} =0, \hspace{1cm} \nabla_{[\mu}F_{\nu\lambda]}=0
\end{equation}
with $\nabla_\mu$ being the covariant derivative with respect to $g_{\mu\nu}$, a Lorentz metric on
$\mathbb{R}^4$. These are 8 equations for the 6 unknowns $F_{\mu\nu}$. Next pick a timelike vector
field $u^\mu$ with $u^2 = -1$ and define electric and magnetic fields by
\begin{equation}\label{EM}
E_\mu = F_{\mu\nu} u^\nu, \hspace{0.8cm}B_{\mu\nu\lambda} = 3 F_{[\mu\nu} u_{\lambda]},
\end{equation}
so that
\begin{equation}\label{EM1}
F_{\mu\nu} =  - 2 E_{[\mu} u_{\nu]} - B_{\mu\nu\lambda} u^\lambda.
\end{equation}
We assume for simplicity that $u^\mu$ is covariant constant, otherwise the ensuing equations
contain zero'th order terms which are of no concern to us. The operator
\begin{equation}\label{der}
\nabla_{\mu\nu} = 2 u_{[\mu} \nabla_{\nu]}
\end{equation}
contains derivatives only in directions orthogonal to $u^\mu$. Using Eq.'s (\ref{max}) we find the
evolution equations
\begin{equation}\label{t}
3 \nabla_{[\mu\nu} E_{\lambda]} = - u^\rho \nabla_\rho B_{\mu\nu\lambda},\hspace{0.8cm}
\nabla^{\lambda\rho}B_{\nu\lambda\rho} = 2 u^\rho \nabla_\rho E_\nu.
\end{equation}
Taking now $u^\lambda \nabla_\rho$ of Eq.(\ref{EM}), we rewrite the evolution equations in the form
\begin{equation}\label{r}
W_{\mu\nu}^{\mu' \nu' \lambda} \nabla_\lambda F_{\mu' \nu'} = 0.
\end{equation}
Now take the positive definite metric
\begin{equation}\label{po}
w^{\mu\nu} = 2 u^\mu u^\nu + g^{\mu\nu}.
\end{equation}
Consider now the positive definite metric $a^{\mu\nu\lambda\rho} = 2 w^{\rho[\mu}w^{\nu]\lambda}$
on the space of 2-forms and use it to raise indices in $W_{\mu\nu}^{\mu' \nu'\lambda}$: One obtains
quantities $W^{\mu\nu\mu'\nu'\lambda}$ satisfying
\begin{equation}\label{fin}
W^{\mu\nu\mu'\nu'\lambda} = W^{\mu'\nu'\mu\nu\lambda},
\hspace{0.8cm}W^{\mu\nu\mu'\nu'\lambda}u_\lambda \sim a^{\mu\nu\mu'\nu'}.
\end{equation}
Thus the Eq.'s (\ref{t}) are symmetric hyperbolic with respect to $u_\mu$. For the characteristic
polynomial one finds
\begin{equation}\label{maxpol}
P(k) \sim (u^\mu k_\mu)^2 (g^{\rho\nu} k_\rho k_\nu)^2.
\end{equation}

\bigskip
We now turn to 2nd order equations. Let us assume that the quantities $M^{\mu\nu}_{AB}$ of
Eq.(\ref{equgen}) satisfy
\begin{equation}\label{reg}
M^{\mu\nu}_{AB} = M^{\nu\mu}_{BA}.
\end{equation}
This is necessarily the case when Eq.(\ref{equgen}) comes from a variational principle, because
then
\begin{equation}\label{var}
M^{\mu\nu}_{AB} \sim \frac{\partial^2 \mathcal{L}}{(\partial \partial_\mu f^A)( \partial
\partial_\nu f^B)},
\end{equation}
where $\mathcal{L} = \mathcal{L}(x,f,\partial f)$. Of course,
since only the quantities $M^{(\mu\nu)}_{AB}$ enter the
differential equation (\ref{equgen}), one might as well have
assumed the stronger conditions $M^{\mu\nu}_{AB}= M^{\nu\mu}_{AB}
= M^{\mu\nu}_{BA}$. This is not usually done in continuum
mechanics. The reason is that, while the PDE (and hence the
characteristic polynomial) is unaffected by the above
symmetrization, other physical quantities in the theory, like the
stress, depend on the unsymmetrized object - and such objects
typically enter, if not the equation, then the natural boundary
conditions for the equation on the surface say of an elastic body
(see \cite{BW}). A related reason is that the symmetrized object
would hide other symmetries - present in some situations - which
are more fundamental such as invariances under isometries. For
example the object $C_{A(B|D|E)}$, with $C_{ABDE}$ the elasticity
tensor of Eq.(\ref{iso}), is not symmetric in $(AB)$ and $(DE)$.
But the latter symmetry is important for understanding the
solutions of the linearized equations of motion when the spacetime
has Killing vectors. The work \cite{Ch1} also uses the
unsymmetrized form of $M^{\mu\nu}_{AB}$, the reason being that in
this approach one is only interested in properties of
$M^{\mu\nu}_{AB}$ which do not change when a total divergence
is added to the Lagrangian, and the stronger symmetry, if present,
would in general be destroyed by such an addition. Next it is assumed
that there exists a pair $X^\mu, \xi_\nu$ satisfying
\begin{equation}\label{reg1}
M^{\mu\nu}_{AB}\xi_\mu \xi_\nu \ \textrm{is  negative definite}
\end{equation}
and
\begin{equation}\label{reg2}
M^{\mu\nu}_{AB} (m^A \eta_\mu) (m^B \eta_\nu) > 0 \ \textrm{for all}\ m^A \eta_\mu \neq 0 \
\textrm{with}\ (X,\eta)=0.
\end{equation}
The conditions (\ref{reg1},\ref{reg2}) essentially state that the PDE is the sum of a ``timelike
part" and an ``elliptic part", the latter obeying the Legendre-Hadamard condition of the calculus of
variations \cite{GH}.
 If the equation (\ref{equgen}) has $l=2$ and satisfies
(\ref{reg},\ref{reg1},\ref{reg2}), the system is called regular hyperbolic
with respect to $\xi$. We now
check that every regular hyperbolic system with respect to $\xi$ is weakly hyperbolic with respect to $\xi$. The
characteristic condition reads
\begin{equation}\label{eigen4}
det(M^{\mu\nu}_{AB}(\eta_\mu + \lambda \xi_\mu)(\eta_\nu + \lambda \xi_\nu)) = 0
\end{equation}
The covector $\eta$ in Eq.(\ref{eigen4}) can be decomposed as $\eta = \frac{(X,\eta)}{(X,\xi)} \xi
+ l$ where $l$ satisfies $(X,l)=0$. Thus we can after redefining $\lambda$ assume that $\eta$ in
Eq.(\ref{eigen4}) has $(X,\eta)=0$. Defining $G_{AB}=-M_{AB}(\xi)=-M^{\mu\nu}_{AB}\xi_\mu \xi_\nu$,
$V_{AB}=M_{AB}(\eta)$, $Q_{AB}=M^{\mu\nu}_{(AB)}\xi_\mu\xi_\nu$, consider the eigenvalue problem
\begin{equation}\label{eigen}
\mathcal{D}\hat{f} = \lambda \mathcal{E}\hat{f},
\end{equation}
in
\begin{displaymath}\label{row}
\hat{f}= \left( \begin{array}{c}
u^A \\
v^B
\end{array} \right)
\end{displaymath}
where the quadratic forms $\mathcal{D}$, $\mathcal{E}$ are given by
\begin{displaymath}\label{eigen1}
\mathcal{D} = \left( \begin{array}{cc}
0      &   V_{AB}\\
V_{AB} &  2Q_{AB}
\end{array} \right)
\end{displaymath}
and
\begin{displaymath}\label{eigen2}
\mathcal{E} = \left( \begin{array}{cc}
V_{AB} & 0\\
0     & G_{AB}
\end{array} \right).
\end{displaymath}
Since $\mathscr{E}$ is positive definite, all eigenvalues $\lambda$ are real. But Eq.(\ref{eigen})
for $\hat{f} \neq 0$ is equivalent to
\begin{equation} \label{eigen3}
(- G_{AB} \lambda^2 + 2 Q_{AB} \lambda + V_{AB})v^B = 0,
\end{equation}
for $v^A \neq 0$ which in turn is equivalent to Eq.(\ref{eigen4}).
This proves our assertion that regular hyperbolic systems are
weakly hyperbolic. (Note that every ``timelike vector" $X$ in the
sense of Eq.(\ref{reg2}) is causal, i.e. $(X,\xi)\neq 0$ for all
$\xi \in \Gamma^*(\xi)$, but not conversely.) We can now
come back to Ex.5. The leading-order coefficients
$M^{\mu\nu}_{AB}$ in Eq.(\ref{el}) clearly belong to a regular
hyperbolic system, when we choose the vector $X^\mu \sim u^\mu$.
It then follows from the preceding result
that the polynomial in Eq.(\ref{elpol}) is indeed a hyperbolic polynomial.\\
As with
symmetric hyperbolic systems, it turns out that there is, for regular hyperbolic systems, a local
existence theorem \cite{HKM} along the lines sketched at the beginning of this section. The
appropriate
domain of dependence theorem is proved in \cite{Ch1}.\\
One may ask the question if it is possible to convert a regular
hyperbolic system into an equivalent symmetric hyperbolic one by
introducing first derivatives as additional dependent variables
(at the price of course of having to solve constraints for the
initial data). (This was the approach we originally followed for
elasticity in \cite{BS}, since we were unaware that there was
already an existence theorem which applied, namely \cite{HKM}). If
the condition (\ref{pos}) is provisionally ignored, it turns out
this is possible provided that $M^{\mu\nu}_{AB}$ is of the form of
Eq.(\ref{el}) for some pair $u^\mu$, $\tau_\nu$, i.e. certain
cross-terms vanish \footnote{In \cite{B} I claimed this to be
possible even without these cross-terms vanishing. I now see I
have no proof of this assertion.}. But the positivity condition
(\ref{pos}) will not always be satisfied. (Essentially this
requires the ''rank-one positivity'' condition Eq.(\ref{reg2}) to
be replaced by the stronger rank-two positivity:
$M^{\mu\nu}_{AB}m^A{}_\mu m^B{}_\nu >0$ for all $m^A{}_\mu \neq 0$
with $X^\mu m^A{}_\mu =0$.) In the case of isotropic elasticity it
was shown in \cite{BS} that one can add to $M^{\mu\nu}_{AB}$ a
term of the form $\Lambda^{\mu\nu}_{AB}$, which has the symmetries
$\Lambda^{\mu\nu}_{AB} = \Lambda^{[\mu\nu]}_{[AB]}$, so that both
the field equations and the requirement Eq.(\ref{reg}) remains
unchanged, but at the same time condition (\ref{pos}) is valid.
However it is an algebraic fact
that such a trick does not always work (see \cite{Te,Se}).\\
Finally let us mention the notion of strong hyperbolicity, which is intermediate between weak
hyperbolicity and symmetric or regular hyperbolicity in the first or second order case
respectively. This
notion, which involves the tool of pseudodifferential reduction \cite{T1,T2}, also gives
wellposedness but has greater flexibility, see \cite{NOR} for applications to the
Einstein equations. It would be interesting to see if the chain ``weakly hyperbolic - strongly
hyperbolic - symmetric or regular hyperbolic" has an analogue for PDE's of order greater than 2.

\bigskip
{\bf Acknowledgment}: I am indebted to Domenico Giulini for carefully reading the manuscript and
suggesting several corrections and improvements. I also would like to thank Helmuth Urbantke for
several very helpful conversations and Piotr T. Chru\'sciel and Helmut Rumpf for useful comments.
This work as supported by Fonds zur F\"orderung der Wissenschaftlichen Forschung (Projekt
P16745-N02)

\end{document}